# Abusir: from Pliny the Elder to Google Maps

Amelia Carolina Sparavigna
Dipartimento di Fisica,
Politecnico di Torino, Torino, Italy

Abusir, the House of Osiris, is the name given to an Egyptian necropolis of the Old Kingdom period. This site is a part of a huge area, from Giza to Dahshur, rich of archaeological remains and covered by many pyramids. The paper is reporting concisely some archaeological studies on Abusir. We start from the description given by Pliny the Elder and end proposing the use of Google Maps.

The earlier information on Abusir, a necropolis to the south of Cairo, is coming from the Natural History by Pliny the Elder (23 AD – August 25, 79 AD) [1,2]. He is telling: "We must make some mention, too, however cursorily, of the Pyramids of Egypt, so many idle and frivolous pieces of ostentation of their resources, on the part of the monarchs of that country. Indeed, it is asserted by most persons, that the only motive for constructing them, was either a determination not to leave their treasures to their successors or to rivals that might be plotting to supplant them, or to prevent the lower classes from remaining unoccupied. There was great vanity displayed by these men in constructions of this description, and there are still the remains of many of them in an unfinished state. There is one to be seen in the Nome of Arsinoites [3]; two in that of Memphites, not far from the Labyrinth, of which we shall shortly have to speak; and two in the place where Lake Moeris was excavated, an immense artificial piece of water, cited by the Egyptians among their wondrous and memorable works: the summits of the pyramids, it is said, are to be seen above the water. The other three pyramids, the renown of which has filled the whole earth, and which are conspicuous from every quarter to persons navigating the river, are situate on the African side of it, upon a rocky sterile elevation. They lie between the city of Memphis and what we have mentioned as the Delta, within four miles of the river, and seven miles and a-half from Memphis, near a village known as Busiris, the people of which are in the habit of ascending them."

Busiris, that is Abusir, the House or Temple of Osiris, is the name given to a necropolis, originally from the Old Kingdom period. Abusir is located to the north of the archaeological area of Saqqara, and served as one of the cemeteries for the ancient Memphis. As Pliny is describing, Abusir is a part of a huge area covered by pyramids, ranging from Giza to Dahshur. The archaeological interest for Abusir aroused, as in general for all the ancient Egypt, after the Napoleon's campaign.

The first survey of Abusir was made by John Perring in the 1830s [4]. Then the Abusir necropolis was visited by the German expedition led by Richard Lepsius, who prepared a map of the site (in Fig.1, the first page of his book, Briefe aus Aegypten, Aethiopien und der Halbinsel des Sinai geschrieben in den Jahren 1842-1845). The first large research campaign on Abusir site was carried out at the beginning of the 20th century by the German Oriental Society, led by Ludwig Borchardt.

It was in 1902, during the excavation of a grave that the Timotheus papyrus was found in a coffin (Fig.2). Timotheus of Miletus, c. 446-357 BC, Greek musician and poet, composed poems of mythological and historical character. The papyrus found at Abusir, that contains the poem "The Persians", is probably the oldest known Greek papyrus. After the finding, the poem had been edited in 1903 [7,8].

Actually, the Abusir archaeological site is known for the Abusir Papyri, the largest finding of papyri, dating from the period of the Old Kingdom [9]. Some papyri were discovered in 1893 at Abu Gorab. They are dated to around 24-th century BC, during then the Fifth dynasty of Egypt, and are therefore considered as the oldest surviving papyri to date [9]. A large number of additional fragments were discovered during the following excavations, as for example, those found by a Czech expedition in some cult complexes.

Two Italian scholars, Vito Maragioglio and Celeste Rinaldi, contributed to the researches on Abusir too. They surveyed the site providing rich information on the Abusir pyramids and improving the plans of the monuments [4]. At the same time, the beginning of 1960s, a Czech archaeological expedition started to work in this area. After a geophysical survey of the whole site, the Czech team transferred, in the mid 1970s, its researches to the not yet investigated southern area of the necropolis. The Czech Institute of Egyptology is continuing excavations at Abusir, the team led by Miroslav Bárta [10].

Abusir is shown in Fig.3. The image was obtained, after a suitable processing to enhance details and contrast, from the Google Maps imagery. The satellite survey shows how huge is this archaeological site, with several complexes not yet completely excavated. It seems that there are 14 pyramids at this site [11]. According to Wikipedia item, the quality of these constructions is inferior to those of the Fourth Dynasty. It could be due to a decrease in royal power or to a poorer economy. "They are smaller than their predecessors, and are built of low quality local stone". That is, less resources for ostentation, echoing Pliny. Moreover, the item is telling that all of the major pyramids at Abusir were built as step pyramids. The largest of them, the Pyramid of Neferirkare, was in origin a step pyramid some seventy metres high and then modified in a pyramid with smooth faces, by filling its steps with loose masonry. We can verify this point by means of Google Maps. Let us observe the Neferirkare complex: it is given in Fig.4. The original steps of the pyramid are clearly visible.

To have an idea of a possible use of the Google Maps imagery in archaeological studies, let us consider the Necropolis, the location shown in Fig.3. From these Maps, we can have an image that, at a first glance, seems not so useful (see the upper part of Fig.5). After processing it by means of AstroFracTool, a program based on the fractional gradient calculus, Iris and GIMP [12], we obtain the image in lower panel. What in the upper panel looked as insignificant spots are now ruins, probably tombs, buried in the sand (in Fig.6, a detail of the area is proposed). A last example of image processing is shown in Fig.7, on the unfinished and the Sahure pyramids.

After the proposed examples, it is seems that the Google satellite images, accompanied by a proper processing, can be useful in archaeological survey, as they can be for geophysical or geopolitical researches. The processing reveals many details, that are usually unnoticed in the maps. In the case of Abusir, the ruins that we see after the processing, probably, have been already discovered and studied. It is possible however, that in other cases, the proposed method can helping in improvement of archaeological knowledge.

8. Ulrich von Wilamowitz-Möllendorff, Die Perser, aus einem Papyrus von Abusir im Auftrage der Deutschen Orientgesellschaft, Leipzig, J.C. Hinrichs Editor, 1903.
9. Abusir Papyri, http://en.wikipedia.org/wiki/Abusir_Papyri
10. Czech Institute of Egyptology, Charles University in Prague, http://egyptologie.ff.cuni.cz/?req=doc:abusir&lang=en&
11. Abusir, http://en.wikipedia.org/wiki/Abusir
12. Amelia Carolina Sparavigna, Enhancing the Google imagery using a wavelet filter, http://arxiv.org/abs/1009.1590

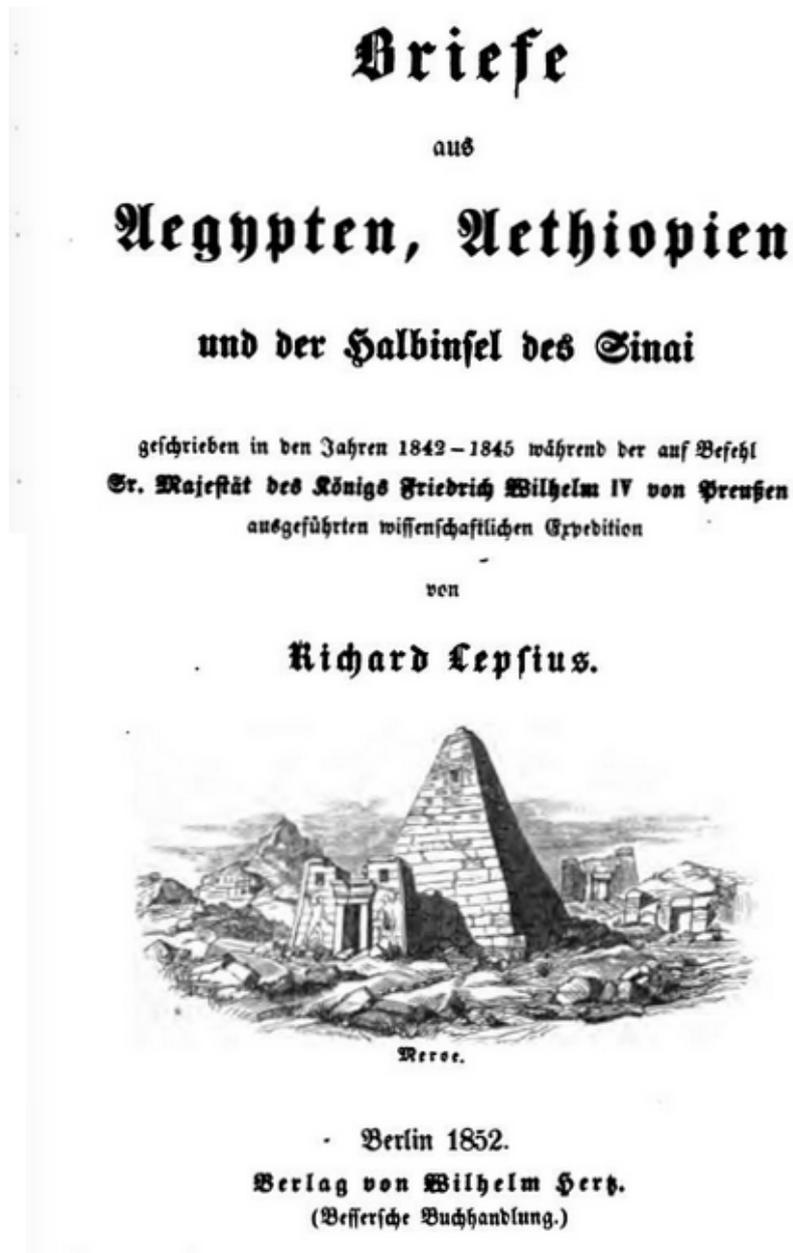

Fig.1: The first page of the book entitled "Briefe aus Aegypten, Aethiopien und der Halbinsel des Sinai geschrieben in den Jahren 1842-1845", by Richard Lepsius. He prepared the first map of the Abusir site.

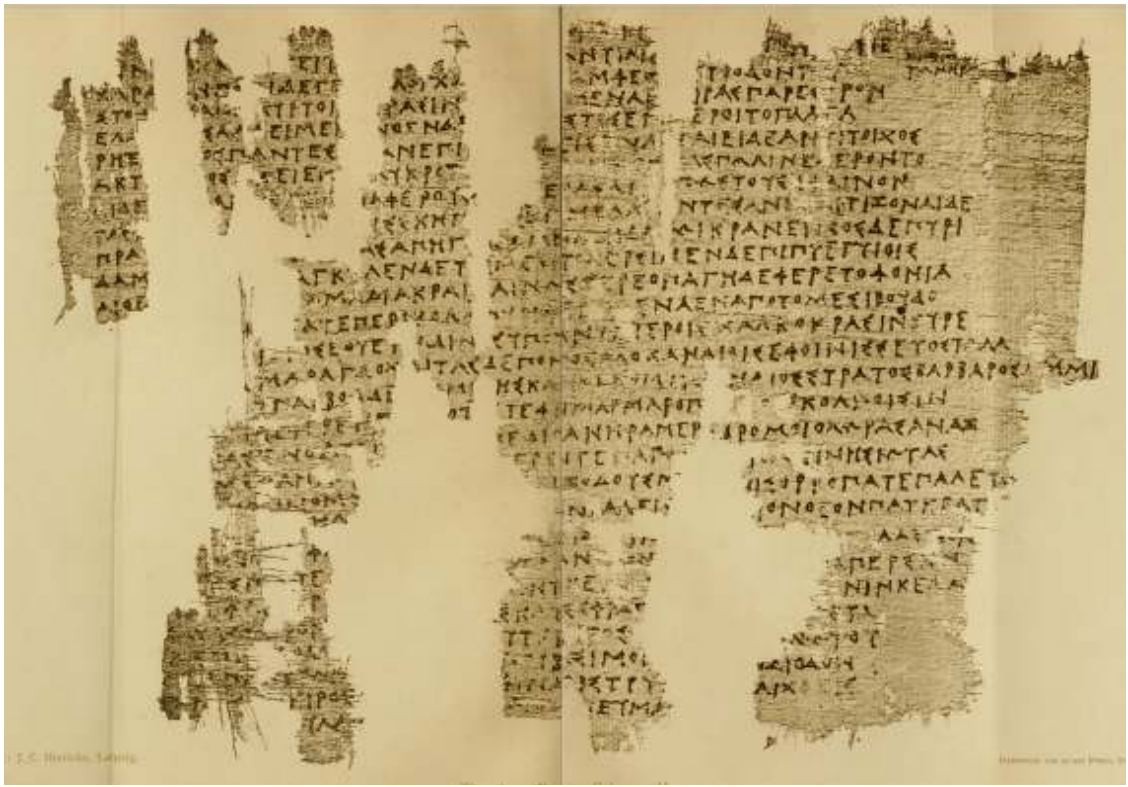

Fig.2: It was in 1902, during the excavation of a grave at Abusir that the Timotheus papyrus was found in a coffin. Timotheus of Miletus (c. 446-357 BC) was a Greek musician and poet. He composed poems of a mythological and historical character. The papyrus found at Abusir is probably the oldest existing Greek papyrus.

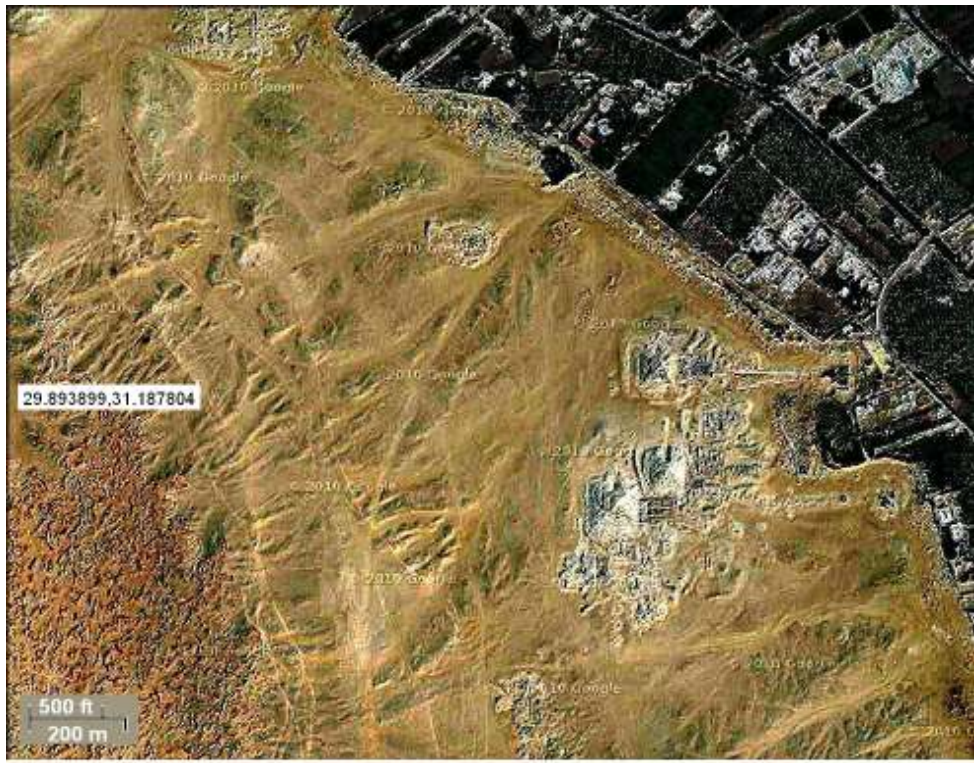
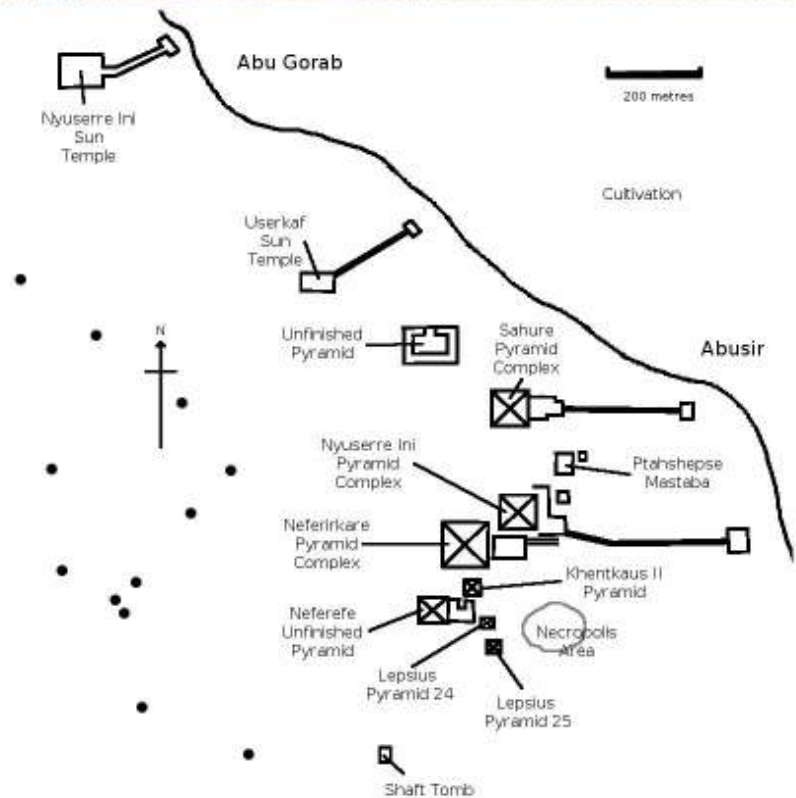

Fig.3: In the upper part of the figure, we see an image from Google Maps, obtained after a suitable image processing, showing the Abusir site. In the lower part, a map from [11], for comparison.

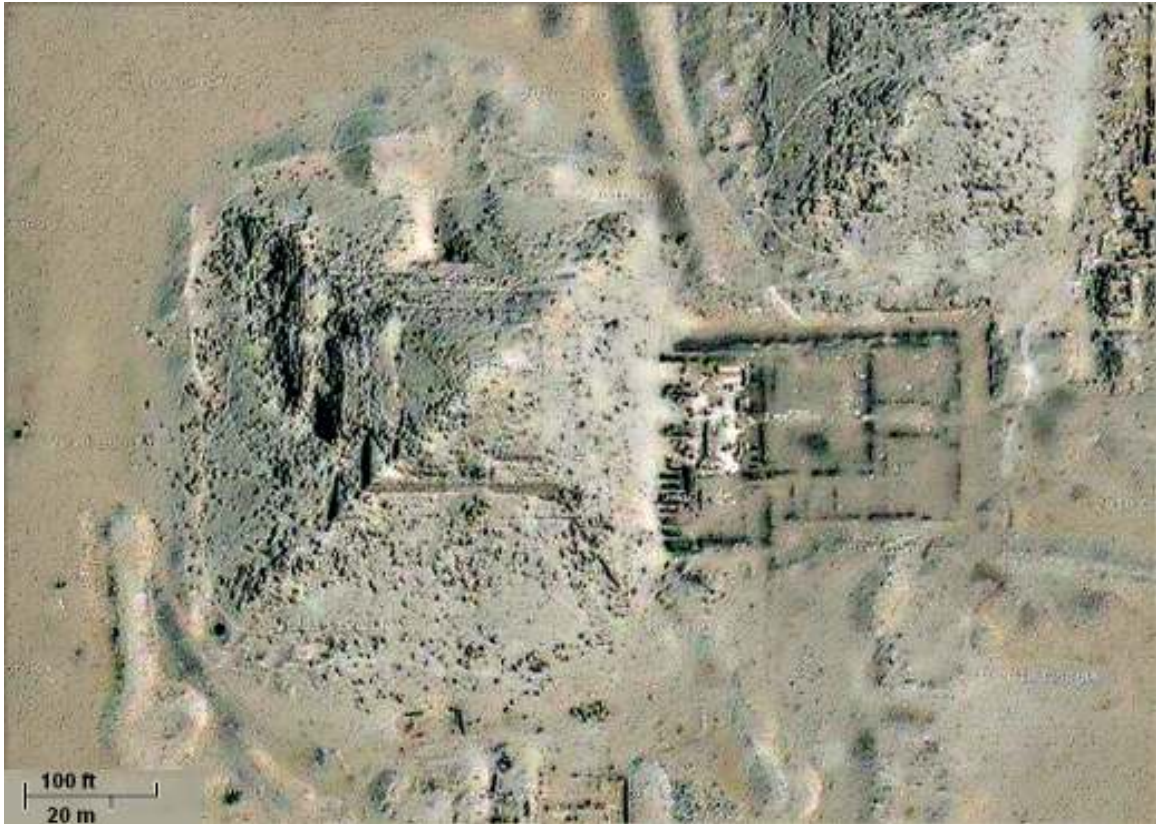

Fig.4: The Pyramid of Neferirkare is the largest pyramid at Abusir. Originally it had been built as a step pyramid and then changed in a smooth pyramid by filling its steps with loose masonry. The image shows as it is seen by means of an image-processing of Google Maps. The original steps of the monument are clearly visible.

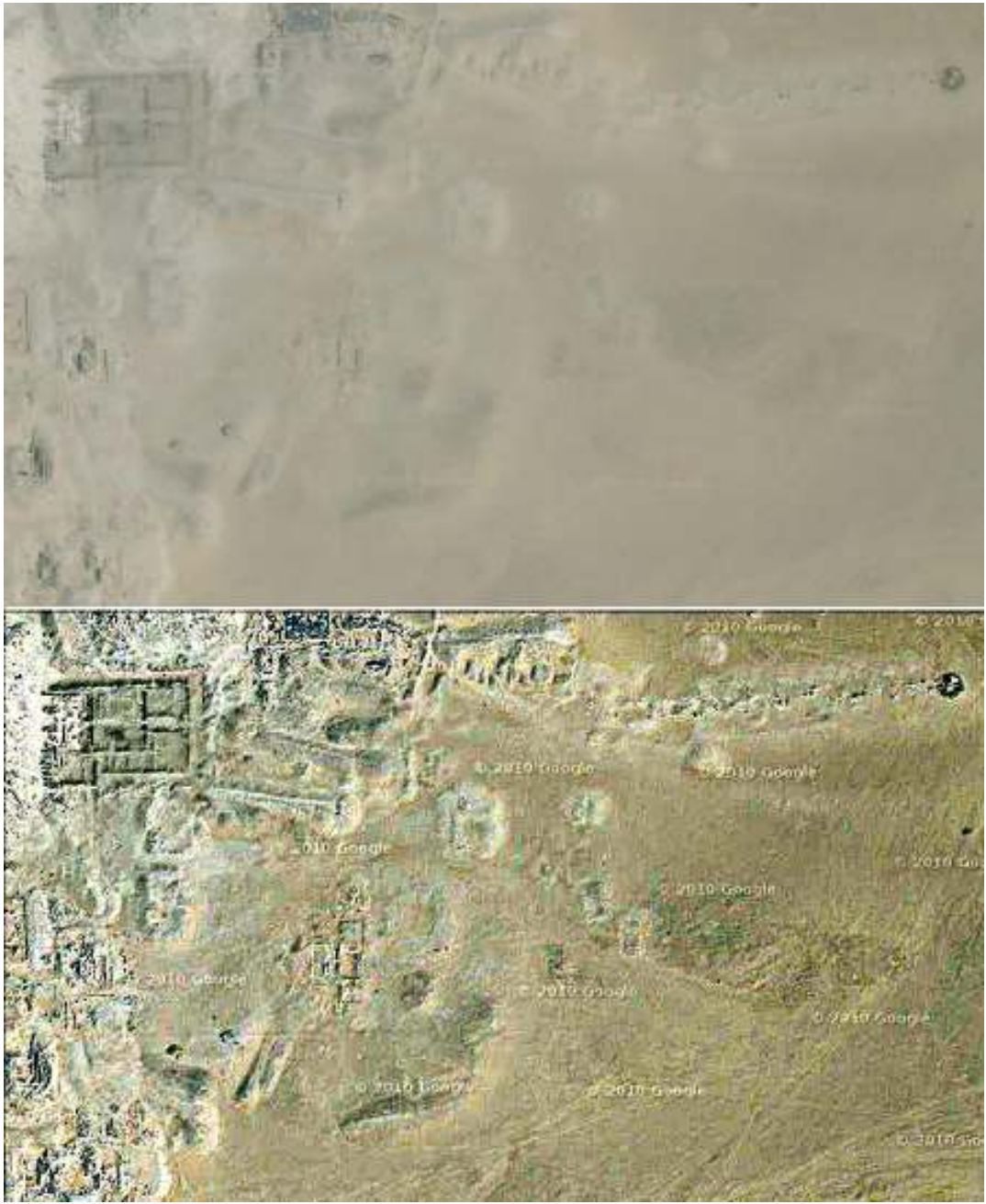

Fig.5: To have an idea of a possible use of image processing in an archaeological survey, let us consider the Necropolis. From the Google Maps, we have an image that, at a first glance, seems not so useful (upper part). After processing it by means of AstroFracTool, a program based on the fractional gradient calculus, Iris and GIMP, we have the image in lower panel. What previously looked as insignificant spots are clearly buildings, probably tombs, buried in the sand

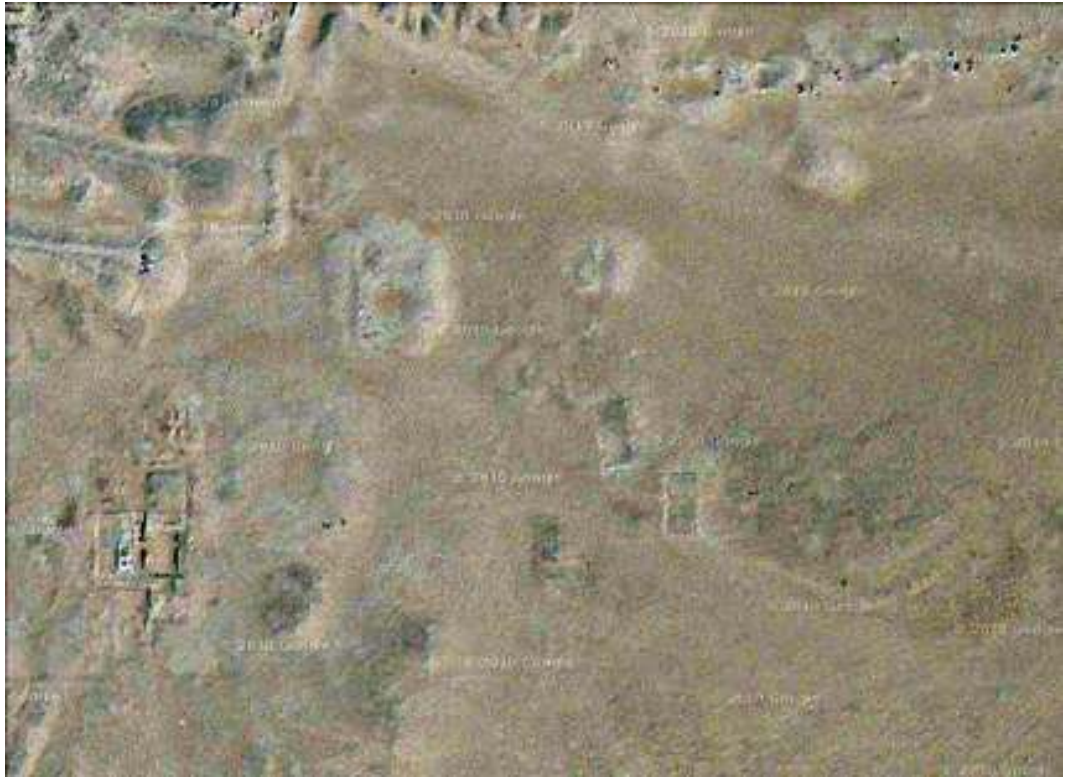

Fig.6: A detail of the Necropolis. The image is obtained from the Google Maps, after image processing with AstroFracTool, Iris and GIMP.

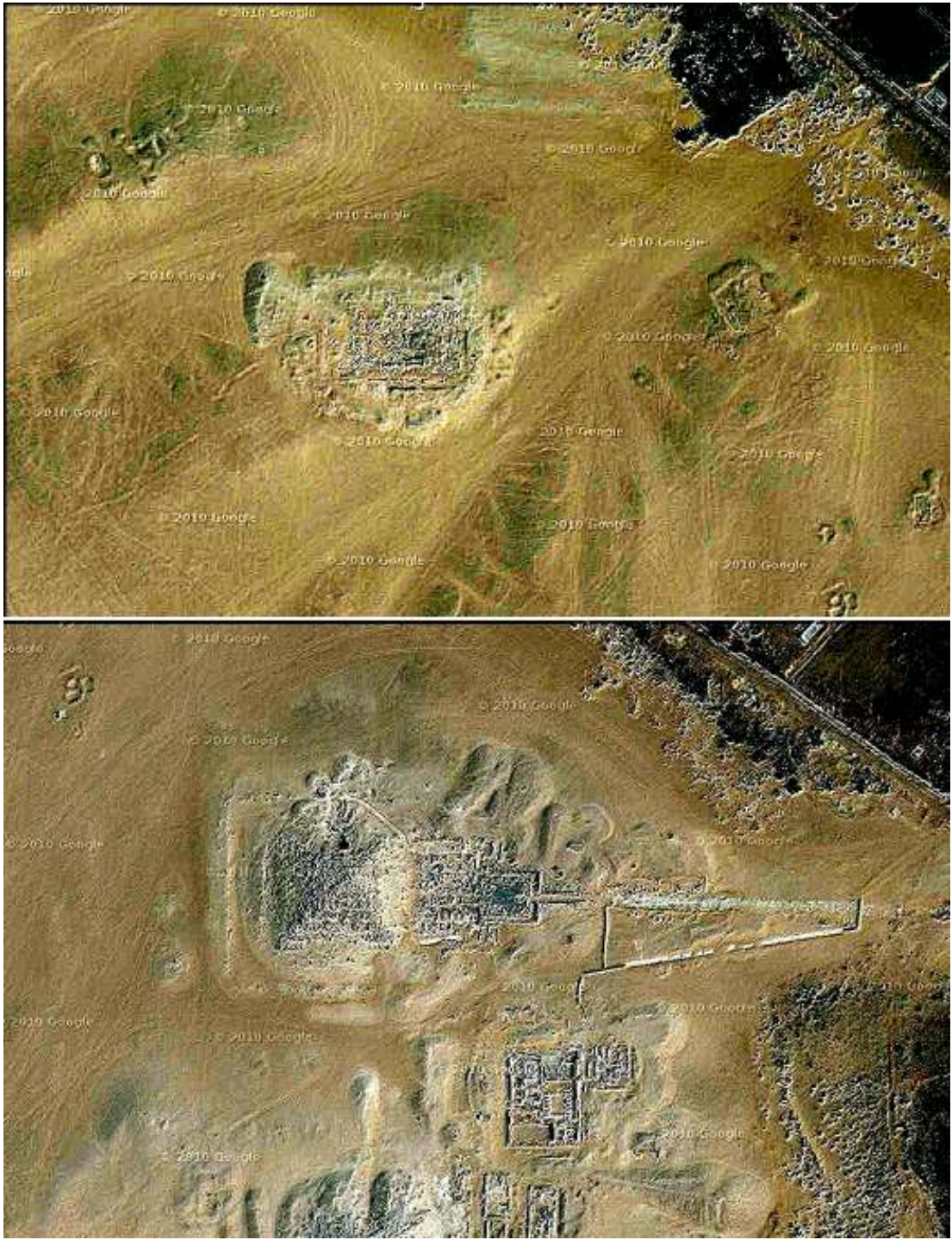

Fig.7: Another example of image processing of Google Maps: in the upper panel the unfinished pyramid, and in the lower panel the Sahure pyramid.